\documentclass[conference]{IEEEtran}
\IEEEoverridecommandlockouts

\usepackage{amsfonts} 
\usepackage{amsmath} 
\DeclareMathOperator*{\argmax}{arg\,max} 
\newcommand{\quotes}[1]{``#1''} 
\usepackage{multirow}

\usepackage{cite}
\usepackage{amsmath,amssymb,amsfonts}
\usepackage{algorithmic}
\usepackage{graphicx}
\usepackage{textcomp}
\usepackage{xcolor}
\def\BibTeX{{\rm B\kern-.05em{\sc i\kern-.025em b}\kern-.08em
    T\kern-.1667em\lower.7ex\hbox{E}\kern-.125emX}}
\begin{document}

\title{Trajectory-User Linking Is Easier Than You Think}

\author{
\IEEEauthorblockN{
Alameen Najjar\IEEEauthorrefmark{1} and
Kyle Mede\IEEEauthorrefmark{2}
\IEEEauthorblockA{
\textit{Rakuten Institute of Technology}, Tokyo, Japan\\
Email: \IEEEauthorrefmark{1}alameen.najjar@rakuten.com, \IEEEauthorrefmark{2}kyle.mede@rakuten.com
}
}
}

\IEEEoverridecommandlockouts
\IEEEpubid{\makebox[\columnwidth]{\hfill}
\hspace{\columnsep}\makebox[\columnwidth]{ }}
\maketitle
\IEEEpubidadjcol

\begin{abstract}
Trajectory-User Linking (TUL) is a relatively new mobility classification task in which anonymous trajectories are linked to the users who generated them. With applications ranging from personalized recommendations to criminal activity detection, TUL has received increasing attention over the past five years. While research has focused mainly on learning deep representations that capture complex spatio-temporal mobility patterns unique to individual users, we demonstrate that visit patterns are highly unique among users and thus simple heuristics applied directly to the raw data are sufficient to solve TUL. More specifically, we demonstrate that a single check-in per trajectory is enough to correctly predict the identity of the user up to $85\%$ of the time. Moreover, by using a non-parametric classifier, we scale up TUL to over 100k users which is an increase over state-of-the-art by three orders of magnitude. Extensive empirical analysis on four real-world datasets (Brightkite, Foursquare, Gowalla and Weeplaces) compares our findings to state-of-the-art results, and more importantly validates our claim that TUL is easier than commonly believed.
\end{abstract}


\section{Introduction}
Trajectory-User Linking (TUL) is a mobility classification task recently introduced by \cite{gao2017identifying} in which anonymous trajectories are linked to the users who generated them. TUL has been claimed essential for a variety of applications, such as mobility data aggregation \cite{sun2021trajectory}, personalized recommendations~\cite{gao2017identifying, zhou2018trajectory, may2020marc, miao2020trajectory}, consumer targeting \cite{gao2020adversarial}, anomaly detection \cite{yu2020tulsn, may2020marc}, criminal/terrorist behavior detection \cite{gao2017identifying, zhou2018trajectory, may2020marc, gao2020adversarial, miao2020trajectory, chen2022mutual}, and epidemic prevention \cite{chen2022mutual}. More importantly, TUL has been used in \cite{rao2020lstm} to quantify the performance of privacy-preserving mobility trajectory synthesis algorithms. In other words, TUL has received an increasing attention from the research community since its introduction in 2017.

Previous works on TUL \cite{gao2017identifying, zhou2018trajectory, may2020marc, gao2020adversarial, miao2020trajectory, yu2020tulsn, sun2021trajectory, zhou2021self, chen2022mutual} have been mainly focused on learning end-to-end deep representations that capture complex spatio-temporal mobility patterns unique to individual users. The learned representations are used to train a classifier that assigns an input trajectory with a label indicating the identity of the user it belongs to. With a maximum number of 800 targeted users \cite{chen2022mutual}, scalability is one major limitation of existing works on TUL. More importantly, previous works have ignored two hallmarks of human mobility data: 1) the uniqueness of mobility patterns among users \cite{de2013unique}, and 2) the high predictability of human mobility patterns over long enough periods of time as shown in \cite{brockmann2006scaling, gonzalez2008understanding,song2010limits, wang2011human, sadilek2012far, krumme2013predictability, hasan2013spatiotemporal}.

In this paper, we argue that since mobility patterns are highly unique among users \cite{de2013unique}, TUL can be solved by simple heuristics applied directly to the raw trajectory data itself. More specifically, we demonstrate that a single check-in per trajectory is enough to correctly predict the identity of the user up to $85\%$ of the time. We validate our argument by conducting an extensive empirical analysis on four real-world check-in datasets widely used for TUL. Moreover, using a non-parametric classifier we scale up TUL to 2.7M trajectories belonging to over 100k unique users which is an increase over state-of-the-art by three orders of magnitude. Contributions made in this paper are summarized as follows:
\begin{itemize}
    \item Empirically demonstrating that check-in patterns are highly unique among users. More specifically, a single check-in per trajectory is enough to correctly predict the identity of the user up to $85\%$ of the time.
    \item Empirically demonstrating that simple heuristics applied to raw trajectory data are not only sufficient to solve TUL but also outperform state-of-the-art approaches.
    \item Scaling up TUL to over 100k users which is an increase over state-of-the-art by three orders of magnitude.
\end{itemize}

The remainder of this paper is organized as follows: Previous works on TUL are reviewed first. Our heuristics-based approach to solving TUL is explained next. Experiments validating our claims are given later. And finally, a summary and discussion concludes the paper.

\section{Previous Works}
TUL was introduced in \cite{gao2017identifying} where sub-trajectories are first embedded into a low dimensional space similar to Word2Vec~\cite{mikolov2013efficient}. Then, fed to a recurrent neural network (RNN)~\cite{rumelhart1985learning} that learns to capture spatio-temporal patterns unique to individual users. The learned representations are finally fed to a Softmax layer with the number of outputs equal to the number of target users. The network is trained end to end, during which inference trajectories are assigned the label with the highest probability. In \cite{zhou2018trajectory} it was shown that Variational Autoencoders (VAE) \cite{kingma2013auto} coupled with RNNs of varying depths capture hierarchical semantics of user trajectories which is reflected in a slightly improved performance. In \cite{may2020marc} a significant improvement over state-of-the-art is reported using a simple three-step architecture. In \cite{gao2020adversarial} Adversarial Generative Networks (GANs) \cite{goodfellow2014generative} coupled with attention mechanism are used to capture complex patterns of user behavior. In \cite{miao2020trajectory} a historical-attention layer is incorporated in RNNs to learn higher-order and multi-periodic patterns. In~\cite{yu2020tulsn} a Siamese network \cite{bromley1993signature} with an RNN backbone is trained to identify trajectories of similar users. The learned representations are used to classify trajectories using a $k$-Nearest Neighbors ($k$-NN) \cite{fix1989discriminatory} classifier. In \cite{sun2021trajectory} an RNN with attention mechanism followed by a multilayer perceptron (MLP) \cite{rosenblatt1958perceptron} with a Softmax classifier are used to link trajectories to users. In \cite{zhou2021self} Contrastive learning is used to solve TUL. Finally, in \cite{chen2022mutual} a mutual distillation learning framework is proposed to learn representations that capture rich contextual check-in patterns using RNNs and temporal-aware transformers.

Previous works have been limited to classifying trajectories of a few hundred users at best. The reason lies in the use of a classification layer with hardcoded outputs one per target user. Training a Softmax layer with a large number of outputs is challenging and computationally expensive \cite{joulin2017efficient}. Not to mention that the model needs to be retrained from scratch every time a new user is added. This makes most of the existing works impractical especially in real-world scenarios where the number of users is ever-increasing. It is worth noting that \cite{yu2020tulsn, zhou2021self} do not explicitly train a classification layer with  hardcoded outputs and thus they are potentially scalable. However, scalability was not investigated in either study.

\section{Trajectory-User Linking}
In this section we first define TUL as a problem. Then, we proceed to explain our approach to solving it.

\subsection{Problem definition}
Let the triplet $(u,t,v)$ denote a record of user $u$ checking-in at venue $v$ and time $t$. The sequence of $n$ chronologically ordered check-in records generated by user $u$ over time interval $\tau$ is called a trajectory, given as $T_{u\tau} = \{(u,t_1,v_1), (u,t_2,v_2)\},\cdot\cdot\cdot,(u,t_n,v_n)\}$. Solving TUL translates to finding a mapping $\mathcal{T}\mapsto\mathcal{U}$, where $\mathcal{T} = \{T_{1\tau}, T_{2\tau},\cdot\cdot\cdot, T_{m\tau}\}$ is the set of $m$ trajectories generated by $m$ users given as $\mathcal{U} = \{u_1, u_2,\cdot\cdot\cdot, u_m\}$. 

The following explains how we solve TUL in three successive steps: 1) trajectory segmentation, 2) trajectory encoding, and finally 3) trajectory classification.

\subsection{Trajectory segmentation}
To reduce computational complexity \cite{gao2017identifying} and capture meaningful temporal patterns we segment each trajectory into $k$ consecutive sub-trajectories spanning a shorter time interval~$\tau$. The segmented trajectory is given as $T_{u\tau} = \{T_{u_1}, T_{u_2}, \cdot\cdot\cdot, T_{u_k}\}$. Segmentation interval $\tau$ can be set to any time period, such as a day or an hour. In our case we use three time intervals, namely day, week and month.



\subsection{Trajectory encoding}
Since sub-trajectories vary in length (Number of check-ins), first we need to transform the input sub-trajectory into a vector $r_{u_i}$ of a unified dimension $d$ such that $r_{u_i} \in \mathbb{R}^d$. Such a vector can be learned in an unsupervised fashion via projecting the check-ins into a space of dimension $d$. However, for reasons that will be clear in the following section, we obtain this representation by sampling $d$ values from the input sub-trajectory, such that:

\begin{equation}
    r_{u_i} = (v_1, v_2, \cdot\cdot\cdot, v_d) \in \mathbb{R}^d,
\end{equation}
where $v_i \in \mathbb{R}$ is the ID of a venue visited by user $u$ and sampled from a set $V_d$ given by:

\begin{equation}
    V_d = \argmax_{\hat{V} \subset V, |\hat{V}|=d} \sum_{v \in \hat{V}} v
\end{equation}

In other words, a sub-trajectory $T_{u_i}$ of a user $u$ is encoded by a concatenation of its $d$ largest venue IDs. This vector is used next for classification.

\subsection{Trajectory classification}
To classify trajectories we use a $k$-Nearest Neighbors classifier. $k$-NN is a non-parametric supervised classification algorithm in which an unlabeled object is assigned a class label based on the majority class membership of its $k$ nearest neighbors. The neighbors are drawn from a set of objects of which the labels are known. This can be thought of as the training dataset for the algorithm, though no explicit training step is required. Proximity among objects is determined using a distance metric, such as the Euclidean distance:

\begin{equation}
    d(x, \hat{x}) = \sqrt{(x_1-\hat{x_1})^2+\cdot\cdot\cdot+(x_d-\hat{x_d})^2},
\end{equation}
where $x$ and $\hat{x}$ are two $d$-dimensional vectors. The input $x$ is assigned the majority class label of its $k$ closest neighbors in the Euclidean space, such that:

\begin{equation}
    P(y=j|X=x) = \frac{1}{k} \sum_{i \in A} I(y^{(i)}=j),
\end{equation}
where $k$ is the number of the nearest neighbors drawn from the training set $A$ and $I(x)$ is the membership function that evaluates to 1 when the argument $x$ is true and 0 otherwise. It is worth noting that $k$ is the only hyperparameter of the $k$-NN algorithm and it is usually set to an odd value to prevent tie situations.

In our case, the $k$-NN classifier takes a $d$-dimensional encoding of the trajectory as an input and produces a label indicating the user identity as an output.

\section{Experiments}
In this section we present the results of validating our approach to TUL.

\subsection{Datasets}
We used four check-in datasets widely used for TUL: Brightkite \cite{cho2011friendship}, Foursquare \cite{yang2016participatory}, Gowalla \cite{cho2011friendship} and Weeplaces~\cite{weeplaces}. The datasets are summarized in Table~\ref{tab1}. 

\begin{table}[h!]
  \caption{Summary of the four datasets we used. \quotes{Venues} and \quotes{Users} indicate the number of unique venues and unique users per dataset, respectively. \quotes{Date range} indicates the dates of the earliest and the latest check-ins per dataset.}
  \label{tab1}
  \begin{tabular}{|c|cccc|}
    \hline
    Dataset    & Check-ins & Venues & Users & Date range \\
    \hline
    Brightkite & 4,747,281  & 772,966 & 51,406 & 3/2008 - 10/2010 \\ 
    
    Foursquare & 33,263,631  & 3,680,126 & 266,909 & 4/2012 - 9/2013 \\ 
    
    Gowalla    & 6,442,892  & 1,280,969 & 107,092 & 2/2009 - 10/2010 \\ 
    
    Weeplaces  & 7,369,712  & 971,307 & 15,793 & 11/2003 - 6/2011 \\ 
    \hline
\end{tabular}
\end{table}
\begin{table}[h!]
  \caption{Summary of the twelve datasets we generated. B, F, G, and W stands for Brightkite, foursquare, Gowalla and Weeplaces, respectively.}
  \label{tab2}
  \begin{tabular}{|c|ccccc|}
    \hline
            & Interval & Check-ins & Venues & Trajectories & Users \\
    \hline
            & Daily & 2,927,015 & 498,747 & 409,474 & 8741 \\  
 B & Weekly & 3,374,309  & 464,840 & 199,194 & 6781 \\ 
            & Monthly & 2,483,625 & 350,171 & 44,393 & 2986 \\ 
 \hline
            & Daily & 12,602,613 & 2,087,220 & 2,724,388 & 102,877 \\  
 F & Weekly & 16,506,383 & 2,208,997 & 1,547,091 & 84,344 \\ 
            & Monthly & 7,956,130 & 1,197,817 & 245,966 & 20,220 \\ 
 \hline
            & Daily & 2,997,987 & 830,727 & 481,105 & 17,112 \\ 
 G    & Weekly & 3,082,292 & 762,433 & 196,443 & 10,893 \\ 
            & Monthly & 672,402 & 214,249 & 12,057 & 1086 \\  
 \hline
            & Daily & 5,491,420 & 818,917 & 1,007,607 & 12,759 \\ 
 W  & Weekly & 6,834,597 & 907,490 & 364,035 & 12,119 \\
            & Monthly & 3,885,605 & 547,283 & 57,736 & 4486 \\ 
    \hline
\end{tabular}
\end{table}

We processed each of the datasets at 3 different timescales (Day, week and month) generating 12 datasets in total. We kept daily, weekly and monthly trajectories with a minimum of 3, 5 and 10 check-ins, respectively. Similar to \cite{may2020marc} we only kept users with a minimum of 10 trajectories each. We replaced user IDs and venue IDs with a sequential number starting from 0. And finally, we sorted the datasets by user ID and timestamp in an ascending order. 

\emph{It is worth noting that the preprocessing procedure we follow is similar to what has been adopted in previous works}~\cite{gao2017identifying, zhou2018trajectory, may2020marc, gao2020adversarial, miao2020trajectory, yu2020tulsn, sun2021trajectory, zhou2021self, chen2022mutual}.

The obtained datasets are diverse in terms of timescale, user count, venue count, and number of check-ins. The 12 datasets are summarized in table \ref{tab2}. We used these datasets in all experiments.

\subsection{Evaluation}
For evaluation we used Accuracy at $k$ (ACC@$k$), Macro F1 score (Macro-F1), Macro precision (Macro-P) and Macro recall (Macro-R) as they are the most commonly used evaluation metrics for TUL\cite{gao2017identifying, zhou2018trajectory, may2020marc, gao2020adversarial, miao2020trajectory, yu2020tulsn, sun2021trajectory, zhou2021self, chen2022mutual}. 

ACC@$k$ measures the ratio of correctly classified trajectories among the top $k$ predicted labels:

\begin{equation}
    \text{ACC@}k = \frac{\text{\# of correctly classified trajectories @} k}{\text{\# of all trajectories}}
\end{equation}

Macro-P and Macro-R are the mean precision and recall among all classes, calculated such that:

\begin{equation}
    \text{Macro-P} = \frac{1}{|\mathcal{U}|} \sum_{u \in \mathcal{U}} \frac{\text{TP}_u}{\text{TP}_u + \text{FP}_u}
\end{equation}

\begin{equation}
    \text{Macro-R} = \frac{1}{|\mathcal{U}|} \sum_{u \in \mathcal{U}} \frac{\text{TP}_u}{\text{TP}_u + \text{FN}_u}
\end{equation}
where $u$ is user label, $\text{TP}_u$, $\text{FP}_u$, and $\text{FN}_u$ are the number of true positives, false positives and false negatives for class/user $u$, respectively. 

\begin{figure}[h!]
  \centering
  \includegraphics[width=1\linewidth]{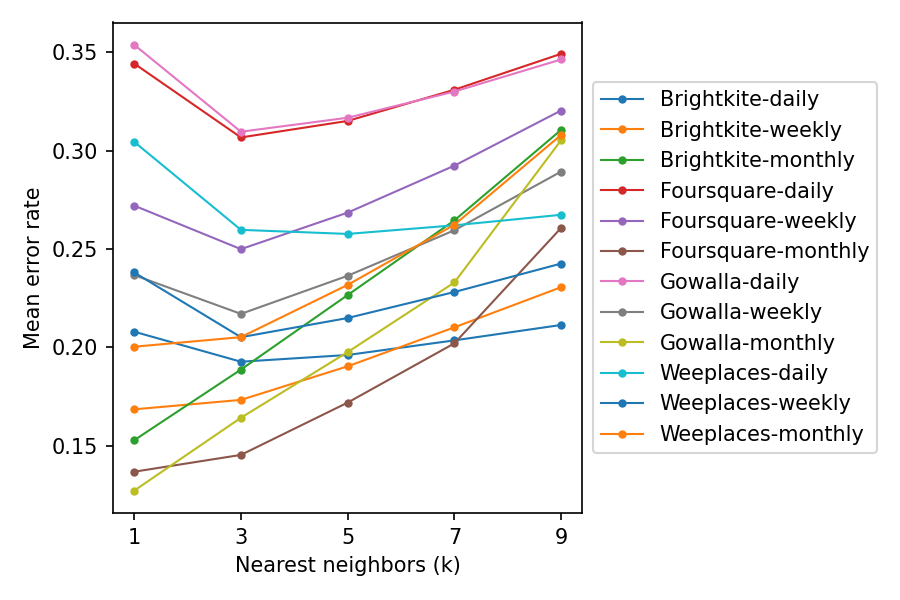}
  \caption{Hyperparameter tuning: Mean error rate plotted against number of nearest neighbors ($k$).}
  \label{figone}
\end{figure}

Finally, Macro-F1 is the harmonic mean of Macro-P and Macro-R averaged over all classes:

\begin{equation}
    \text{Macro-F1} = \frac{2 \cdot \text{Macro-P} \cdot  \text{Macro-R}}{\text{Macro-P} + \text{Macro-R}}
\end{equation}

We report the mean value of the above metrics obtained via conducting a three-fold cross validation \cite{mosteller1968data} implemented with stratified sampling.

\subsection{Hyperparameter tuning}
In Figure \ref{figone}, we plotted the mean error rate as a function of an increasing number of nearest neighbors ($k$). We observed that error rate is the lowest for $k=\{1, 3\}$. Since for $k=1$ classification is sensitive to outliers and noise, we set $k$ to 3 in the following whenever $k$-NN is used.

\subsection{Uniqueness of visit patterns}
Inspired by the findings in \cite{de2013unique}, we set out to answer the following question: How unique are the visit patterns among different users? 

\begin{figure*}[h!]
    \centering
    \includegraphics[width=0.9\textwidth]{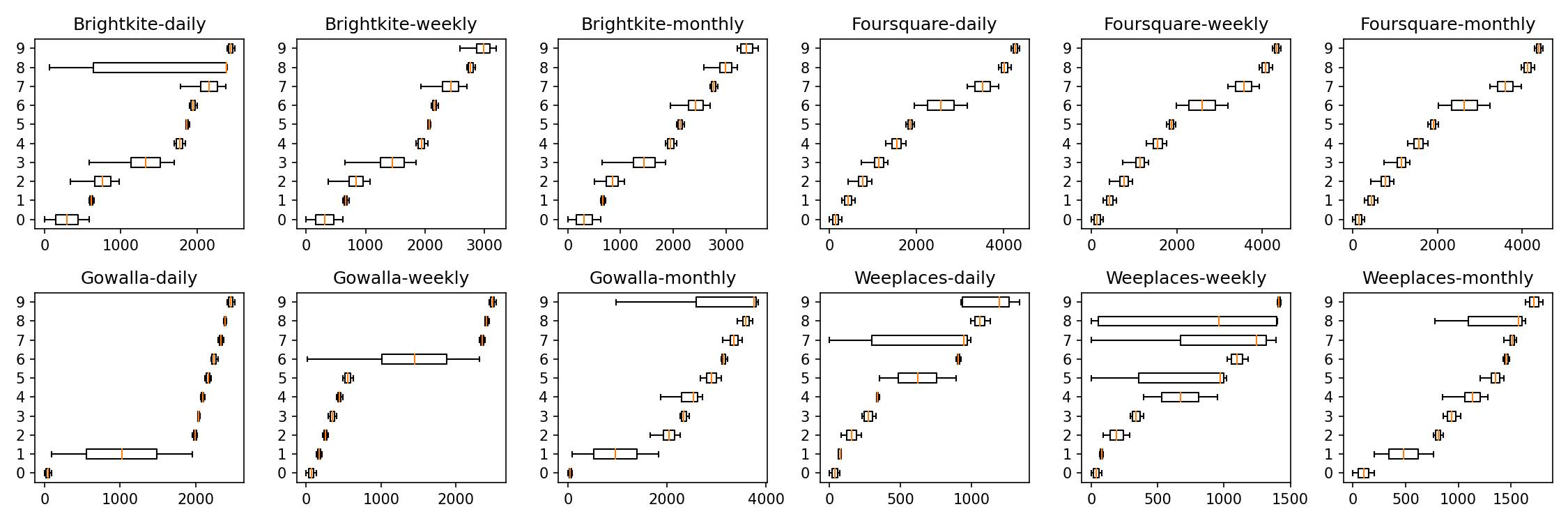} 
    \caption{Distribution of unique venue IDs (X axis) per user ID (Y axis) for the first 10 users in each of our 12 datasets.}
    \label{fig1}
\end{figure*}

\begin{figure*}[h!]
    \centering
    \includegraphics[width=0.9\textwidth]{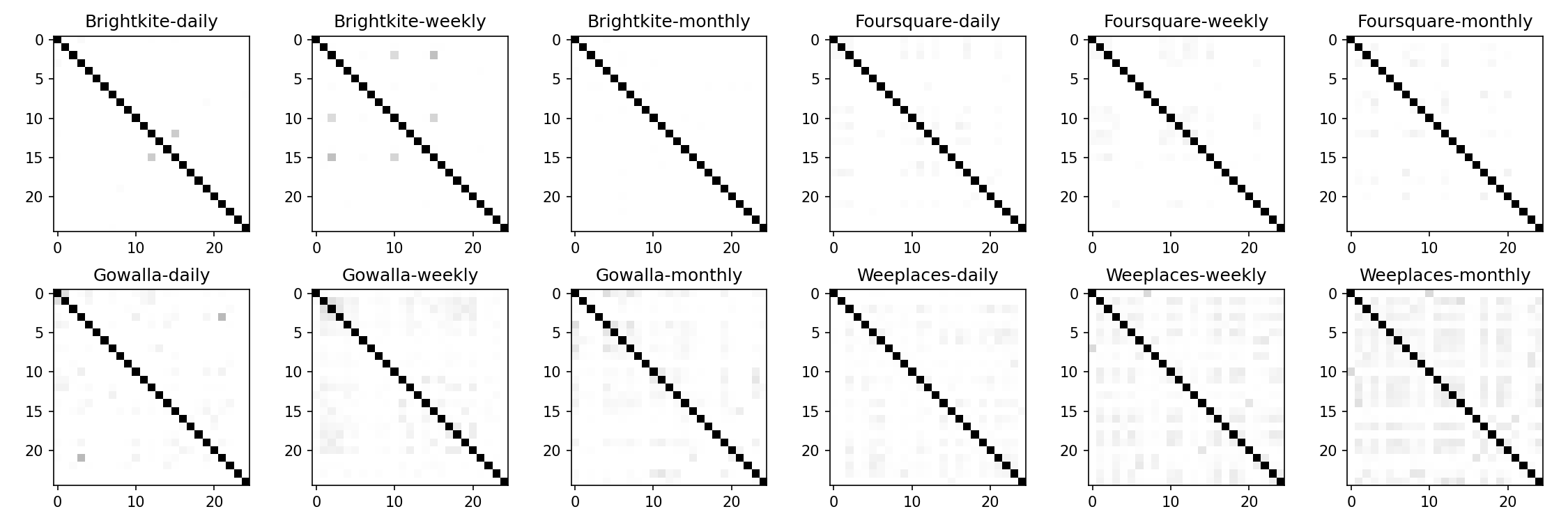} 
    \caption{Jaccard distance matrix calculated between sets of unique venue IDs for the top 25 users in each of the 12 datasets.}
    \label{fig2}
\end{figure*}

We started out by calculating the venue-to-user ratio for all datasets (Table \ref{tab2}). We found that on average there are 47, 60 and 124 unique venues per user for daily, weekly and monthly datasets, respectively. Next, we plotted in Figure \ref{fig1} the distribution of unique venue IDs visited per user for the first 10 users in each of the 12 datasets. From the box plot, it is clear that users are separated in the X axis (Venue ID), i.e., different users visit different venues. We experimented with different user subsets and we obtained similar results. Finally, to empirically quantify the uniqueness of visit patterns among users, we plotted, in Figure \ref{fig2}, the Jaccard distance matrix for the top 25 users in each datasets. It is clear that no two users visit the same set of venues. In fact, the average Jaccard distance among the top 25 users is over 0.996, 0.991 and 0.992 for daily, weekly and monthly trajectories, respectively, i.e., over 99$\%$ of the venues visited by the users are unique to themselves.


The obtained results indicate that visit patterns are highly unique among users which motivates us to investigate whether or not heuristics are sufficient to solve TUL.

\subsection{Are heuristics sufficient to solve TUL?}
Motivated by the above results we evaluate the performance of our simple heuristics-based approach to TUL as introduced in the previous section.

\begin{figure*}[h!]
    \centering
    \includegraphics[width=0.9\textwidth]{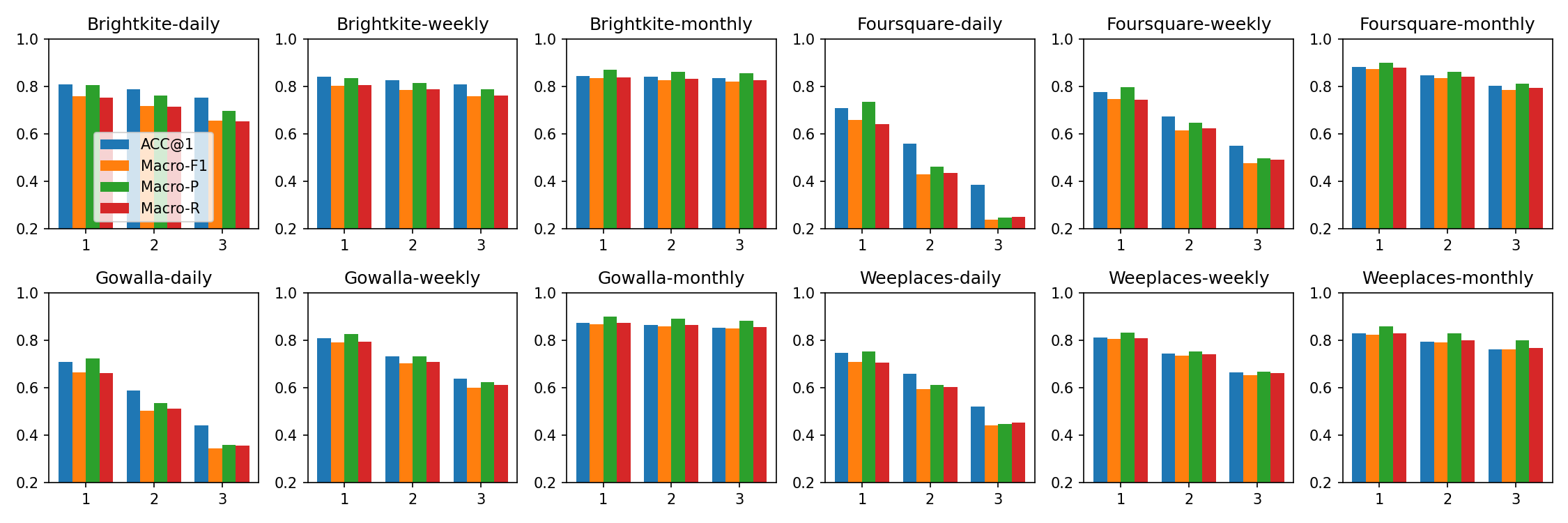} 
    \caption{Classification performance (Y axis) plotted against increasing sampling size $d$ (X axis) on all datasets.}
    \label{fig3}
\end{figure*}

Figure \ref{fig3} summarizes the obtained results plotted as classification performance against increasing value of $d$. We limited~$d$ to 3 since it is the minimum number of check-ins guaranteed per trajectory across different time intervals. On all datasets, smaller $d$ yields better classification results. In fact, for $d=1$, average F1-score is 74.4$\%$, 80.9$\%$, 85.7$\%$ for daily, weekly and monthly trajectories, respectively. This is significant since it means that a single check-in per trajectory is enough to correctly predict the identity of the user up to $85\%$ of the time. Keep in mind that for $d=1$, the Euclidean distance becomes subtraction and consequently the $k$-NN classification resembles a simple thresholding operation applied to the venue~ID.

It is also clear that the degree by which performance degrades with increasing $d$ is reversely related to time interval. Moreover, classification performance degrades the shorter the time interval is. This observation holds true across all datasets. We think that this behavior is likely due to the fact that over long enough periods of time, human mobility patterns are highly predictable \cite{brockmann2006scaling, gonzalez2008understanding,song2010limits, wang2011human, sadilek2012far, krumme2013predictability, hasan2013spatiotemporal}.

Finally, it is worth mentioning that we have experimented with heuristics other than what is proposed in the previous section. More specifically, since as we have demonstrated that different users visit different venues, it is more straightforward to classify trajectories based on how common venues are between trajectories, i.e., replacing the Euclidean distance with the Jaccard distance inside the $k$-NN classifier and using raw trajectories instead of samples. The results we obtained are very similar to what we reported in Figure \ref{fig3} however at the expense of much higher computational cost. This can be attributed to the fact that trajectories can get very long reaching up to a few hundred venues in some cases. Moreover, we experimented with different trajectory sampling methods other than $\max$, such as $\min$ and $median$. However, $\max$ consistently outperformed others which is inherent in the way the data is pre-processed which naturally leads to the $\max$ venue ID having a high likelihood of being a unique identifier for the user as depicted in Figure \ref{fig1}.

\begin{figure*}[h!]
    \centering
    \includegraphics[width=0.9\textwidth]{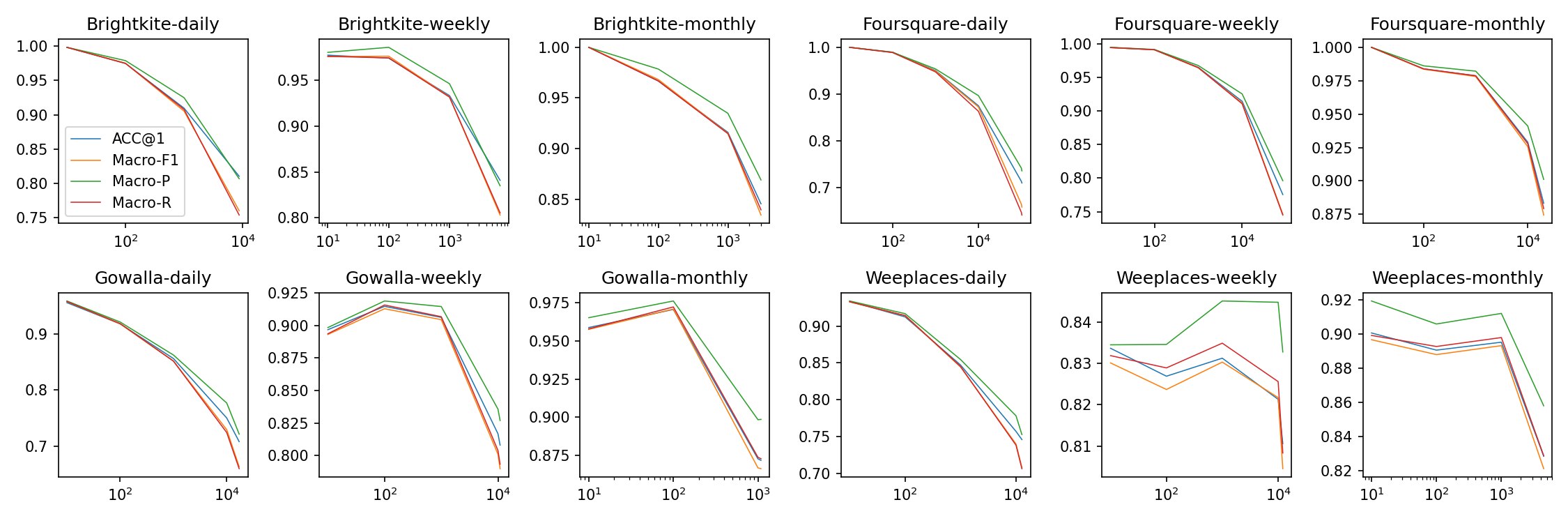} 
    \caption{Scaling up TUL: Classification performance (Y axis) plotted against the number of users (X axis) on all datasets.}
    \label{fig5}
\end{figure*}

In conclusion, the obtained results validate our claim that since venues are highly unique among users, simple heuristics applied to the raw data are sufficient to solve TUL up to $85\%$ of the time.
 
\subsection{Performance vs. time interval}
\begin{figure}[h!]
    \centering
    \includegraphics[width=0.8\columnwidth]{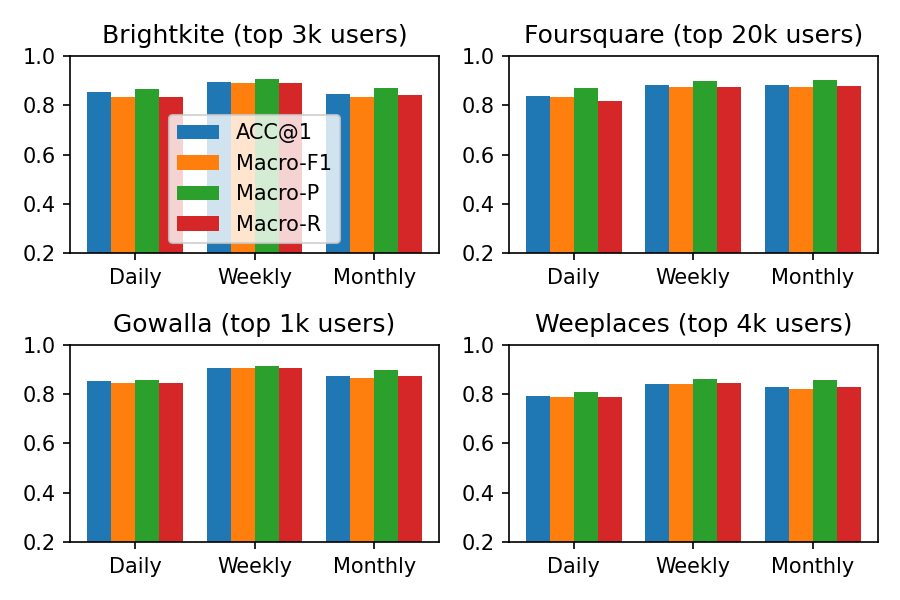}
    \caption{Classification performance (Y axis) as a function of time interval (X axis). Number of users is set to top $N$ where $N$ is the smallest number of users among the three time intervals per dataset.}
    \label{fig4}
\end{figure}

In Figure \ref{fig4}, we plot classification performance against time interval on all datasets. In order to isolate the effect time interval has on performance, we fixed the number of users to that of the minimum among all time intervals per dataset. The obtained results mostly support those of the previous experiment in that performance and time interval length are directly related.

\subsection{Scaling up TUL}
In Figure \ref{fig5} we plotted classification performance against an increasing number of users. User count increases by a single order of magnitude starting from 10. It is worth noting that increasing the number of users implies increasing the number of trajectories and thus increasing the search space for the $k$-NN algorithm. Therefore, it is expected for classification performance to drop as the number of users increases as illustrated in Figure \ref{fig5}.

Moreover, in Table \ref{tab3} we report average classification time on a single machine with 2.4 GHz of processing power and 32 GB of RAM. For the largest dataset (Foursquare daily), on average, classification takes roughly 1.1 millisecond per user (Average of three runs). Keep in mind that time heavily depends on both number of queries and the search space which in this case are $\approx$900k and 1.8M, respectively. 

\renewcommand{\arraystretch}{1.1}
\begin{table*}[h!]
\caption{Comparison with state-of-the-art: Classification performance (ACC@1, ACC@5, Macro-F1, Macro-P, Macro-R) on four datasets. Bold and underline indicate best and second best results, respectively.}
\centering
\resizebox{0.95 \textwidth}{!}{
\begin{tabular}{| c | c | c  c  c  c  c | c  c  c  c  c |}
     \hline
     \multirow{2}{*}{Dataset}  & \multirow{2}{*}{Method}  & ACC@1 & ACC@5 & Macro-F1 & Macro-P & Macro-R & ACC@1 & ACC@5 & Macro-F1 & Macro-P & Macro-R \\
     \cline{3-7} \cline{8-12}
             &         &\multicolumn{5}{c|}{$|\mathcal{U}|=92$} & \multicolumn{5}{c|}{$|\mathcal{U}|=300$} \\
    \hline
     \multirow{9}{*}{Brightkite} &   \cite{gao2017identifying} & 0.4500 & 0.6464 & 0.3938 & - & -  & - & - & - & - & -\\
     & \cite{zhou2018trajectory} & 0.4598 & 0.6484 & 0.4132 & 0.4315 & 0.3965  & - & - & - & - & -\\
     & \cite{yu2020tulsn} & \underline{0.6549} & \underline{0.8275} & \underline{0.6091} & - & - & - & - & - & - & - \\ 
     & \cite{gao2020adversarial} & 0.4891 & 0.6544 & 0.4335 & 0.4677 & 0.4039 & - & - & - & - & -\\
     & \cite{sun2021trajectory} & 0.5845 & 0.7658 & 0.5472 & \underline{0.5656} & \underline{0.5299}  & - & - & - & - & -\\
     & \cite{zhou2021self} & 0.4719 & 0.6498 & 0.4260 & 0.4485 & 0.4055  & - & - & - & - & -\\
     & \cite{may2020marc} & - & - & - & - & -& 0.9492 & \textbf{0.9761} & 0.9304 & 0.9390 & 0.9319 \\ 
     & Ours (Daily) & \textbf{0.9758} & \textbf{0.9808} & \textbf{0.9763} & \textbf{0.9801} & \textbf{0.9759}  & - & - & - & - & -\\ 
     & Ours (Weekly) & - & - & - & - & -  & \textbf{0.9546} & 0.9706 & \textbf{0.9568} & \textbf{0.9715} & \textbf{0.9546}\\ 
     \hline
    \multirow{5}{*}{Foursquare} &        & \multicolumn{5}{c|}{$|\mathcal{U}|=209$} & \multicolumn{5}{c|}{$|\mathcal{U}|=498$} \\
    \cline{3-12}
    & \cite{miao2020trajectory} & 0.6306 & 0.7557 & 0.5024 & 0.5266 & 0.4803 & - & - & - & - & -\\
    & Ours (Daily) & \textbf{0.9809} & \textbf{0.9877} & \textbf{0.9800} & \textbf{0.9812} & \textbf{0.9798} & - & - & - & - & - \\ 
     & \cite{may2020marc} & - & - & - & - & - & 0.9667 & \textbf{0.9887} & 0.9628 & 0.9667 & 0.9624\\ 
     & Ours (Weekly) & - & - & - & - & - & \textbf{0.9751} & 0.9837 & \textbf{0.9748} & \textbf{0.9765} & \textbf{0.9749}\\ 
    \cline{3-12}
     \hline
    \multirow{10}{*}{Gowalla} &        & \multicolumn{5}{c|}{$|\mathcal{U}|=201$} & \multicolumn{5}{c|}{$|\mathcal{U}|=300$} \\
    \cline{3-12}
     & \cite{gao2017identifying} & 0.4570 & 0.6568 & 0.3577 & - & - & - & - & - & - & -\\ 
     & \cite{zhou2018trajectory} & 0.4540 & 0.6239 & 0.3541 & 0.3613 & 0.3471 & - & - & - & - & - \\
     & \cite{yu2020tulsn} & \underline{0.7510} & \underline{0.8997} & \underline{0.7113} & - & - & - & - & - & - & - \\ 
     & \cite{gao2020adversarial} & 0.4761 & 0.6464 & 0.3774 & 0.3995 & 0.3577 & - & - & - & - & - \\ 
     & \cite{sun2021trajectory} & 0.4884 & 0.6682 & 0.4001 & \underline{0.4038} & \underline{0.3964} & - & - & - & - & - \\ 
     & \cite{zhou2021self} & 0.4571 & 0.6398 & 0.3615 & 0.3647 & 0.3583 & - & - & - & - & - \\ 
     & \cite{may2020marc} & - & - & - & - & - & 0.8831 & \textbf{0.9531} & 0.8681 & 0.8893 & 0.8690 \\ 
     & Ours (Daily) & \textbf{0.8948} & \textbf{0.9266} & \textbf{0.8926} & \textbf{0.8969} & \textbf{0.8935} & - & - & - & - & - \\
     & Ours (Weekly) & - & - & - & - & - & \textbf{0.9160} & 0.9401 & \textbf{0.9154} & \textbf{0.9218} & \textbf{0.9180} \\ 
     \hline
     \multirow{3}{*}{Weeplaces} &        & \multicolumn{5}{c|}{$|\mathcal{U}|=400$} & \multicolumn{5}{c|}{$|\mathcal{U}|=800$} \\
        \cline{3-12}
     & \cite{chen2022mutual} & 0.4531 & 0.5828 & 0.4522 & 0.4981 & 0.4424 & 0.4190 & 0.5551 & 0.4162 & 0.4663 & 0.4058\\ 
     & Ours (Daily) & \textbf{0.8753} & \textbf{0.9028} & \textbf{0.8741} & \textbf{0.8803} & \textbf{0.8741} & \textbf{0.8562} & \textbf{0.8876} & \textbf{0.8543} & \textbf{0.8636} & \textbf{0.8550} \\
    \hline
\end{tabular}
}
\label{tab4}
\end{table*}

\begin{figure}[h!]
  \centering
  \includegraphics[width=0.95\columnwidth]{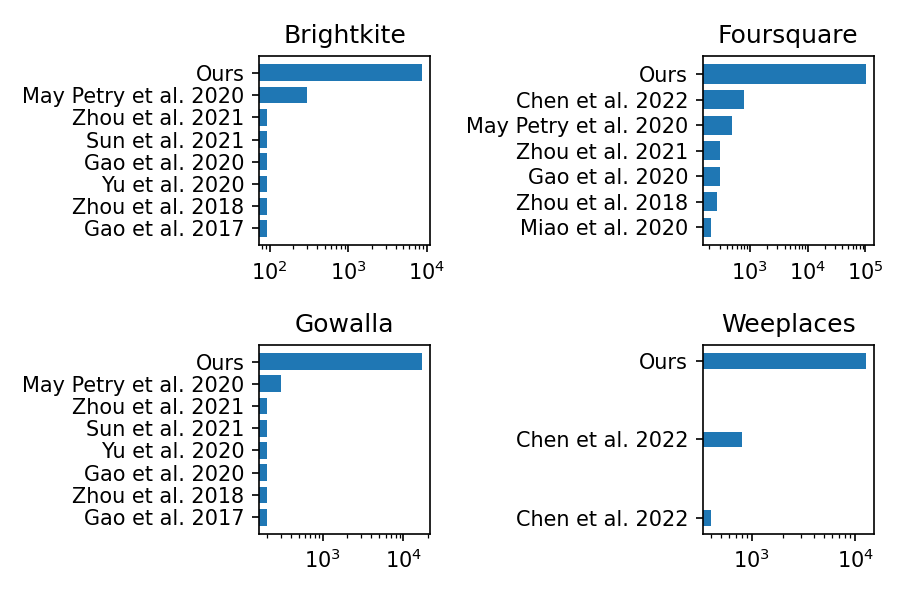}
  \caption{Comparison with existing works: Number of target users (X axis) per dataset (Panels). On Foursquare-daily we classify trajectories of 102,877 users which is over 128 times the number of users reported in the best of state-of-the-art \cite{chen2022mutual}.}
  \label{figx}
\end{figure}

\begin{table}
    \centering
  \caption{Average classification time per user per dataset. \quotes{Queries} and \quotes{Search space} indicate number of test and training trajectories, respectively. Time is reported in milliseconds. B, F, G and W stands for Brightkite, Foursquare, Gowalla and Weeplaces, respectively.}
  \label{tab3}
  \begin{tabular}{|c|ccccc|}
    \hline
            & Interval & Users & Queries & Search space & Time \\
    \hline
            & Daily & 8741 & 164,587 & 334,160 & 0.8 \\  
 B & Weekly & 6781 & 65,734 & 133,460 & 0.44 \\  
            & Monthly & 2986 & 14,650 & 29,743 & 0.01 \\  
 \hline
            & Daily & 102,877 & 899,048 & 1,825,340 & 1.08 \\  
 F & Weekly & 84,344 & 510,540 & 1,036,551 & 0.59 \\  
            & Monthly & 20,220 & 81,169 & 164,797 & 0.2 \\  
 \hline
            & Daily & 17,112 & 158,765 & 322,340 & 0.41 \\  
 G    & Weekly & 10,893 & 64,826 & 131,617 & 0.27 \\  
            & Monthly & 1086 & 3979 & 8078 & 0.01 \\  
 \hline
            & Daily & 12,759 & 332,510 & 675,097 & 1.17 \\  
 W  & Weekly & 12,119 & 120,132 & 243,903 & 0.49 \\  
            & Monthly & 4486 & 19,053 & 38,683 & 0.22 \\ 
    \hline
\end{tabular}
\end{table}

It is worth noting that, to the best of our knowledge, we are the first to scale up TUL to over 100k users. Which is 3 orders of magnitude increase over the best of state-of-the-art as reported in \cite{chen2022mutual}. See Figure \ref{figx} for details.

\subsection{Comparison with state-of-the-art}
For perspective, we compared in Table \ref{tab4} our results to state-of-the-art results on all four datasets. We quoted results as reported in their respective publications. For fair comparison, we used weekly trajectories when comparing with \cite{may2020marc} and daily trajectories otherwise. Moreover, we kept trajectories of the top $|\mathcal{U}|$ users as reported in previous works.

On average our results (Macro-F1) are better than the best of state-of-the-art by 2.6$\%$, 1.2$\%$, 4.7$\%$, and 42$\%$ on Brightkite, Foursquare, Gowalla and Weeplaces, respectively. 

\emph{It is worth noting that we limited the comparison in Table \ref{tab4} to works that used exactly the same datasets we use. And in the case where a dataset has multiple versions (e.g., Foursquare~\cite{yang2016participatory}) we made sure that the same exact version of the dataset is used}.

In summary, the obtained results demonstrate that simple heuristics are not only sufficient to solve TUL but also outperform state-of-the-art works which utilize sophisticated deep learning models.

\subsection{Discussion}
While at first glance it seems counter-intuitive for simple heuristics to outperform deep models, we have clearly demonstrated that visit patterns are highly unique among users,~i.e., different users visit almost completely different venues which is in line with the findings in \cite{de2013unique}, and amplified by the widely adopted TUL evaluation setup where trajectories are drawn from randomly selected users (Most active users usually) using check-in datasets collected from social media accounts of users active in different cities and/or countries. Users in different cities visit completely different venues and thus their trajectories are easily distinguishable using simple heuristics as we demonstrated in the experiments section. 

The question that remains not fully answered is: why is the $\max$ venue ID per sub-trajectory an effective feature? In other words, why is the venue with maximum ID value (Among those visited by the same user in the same sub-trajectory) an effective indicator of the user's identity? The answer simply lies in the way data is preprocessed where check-in records from all users are first anonymized (User ID and venue ID are replaced with sequential numbers). Then, sorted by user ID, venue ID and timestamp. This means that, given a user, venues unique to them (e.g., home) are assigned IDs (Sequential numbers) higher in value than those visited by other users (e.g., restaurant). Therefore, by applying the $\max$ operator to all venue IDs visited by the user in a given sub-trajectory equals to finding the single venue unique to this user among all users. Such a venue could be home (Which is unique to the user unless two or more users share the same house) which is highly likely to be included in most of the sub-trajectories of a given user given long enough time interval. Indeed, this is not always true otherwise our classification results would have been perfect.

We argue that given the current TUL evaluation setup, linking anonymous trajectories to their users is similar to classifying image patches of different color shades. Color shade classification is a simple task that can be done via pixel thresholding without the need for sophisticated vision models with millions of learned parameters. Similarly, since different users visit different venues, solving TUL does not require learning representations that capture complex spatio-temporal mobility patterns unique to individual users. But rather finding the venue(s) unique to each user (e.g., home).

It is worth noting that we are aware of the handful of previous works \cite{zhou2018trajectory, gao2020adversarial, zhou2021self} that experimented with data limited to a single city (Tokyo and/or New York). However, in these same works proposed algorithms were evaluated on other check-in datasets collected from different cities and even countries in some cases.

\section{Summary}
TUL is a relatively new mobility classification task \cite{gao2017identifying} in which anonymous trajectories are linked to the users who generated them. With a variety of proposed applications ranging from personalized recommendations to epidemic prevention and criminal behavior detection, TUL has received increasing attention over the past 5 years. 

Previous works on TUL have been mainly focused on learning deep representations that capture complex spatio-temporal mobility patterns unique to individual users. Scalability is one major limitation of previous works with less than 1000 target users at best.

In this paper we argued that since visit patterns are highly unique among users \cite{de2013unique}, TUL can be solved by simple heuristics applied directly to the raw data. We empirically validated our argument by conducting extensive analysis on four real-world datasets (Brightkite, Foursquare, Gowalla and Weeplaces) and demonstrated that heuristics are not only sufficient to solve TUL but also superior to state-of-the-art approaches which utilize sophisticated models. Moreover, using a non-parametric classifier we scaled up TUL to over 100k users which is an increase over state-of-the-art by three orders of magnitude.

While comparing our results to state-of-the-art is needed for perspective, it is important to remind the reader that we do not aim in this paper to propose a new algorithm to solve TUL but rather invite the research community to rethink existing solutions, datasets and evaluation setup while paying better attention to the statistics and nature of human mobility data in general and social check-in data in specific.

\bibliographystyle{IEEEtran}
\bibliography{bibs}

\begin{thebibliography}{10}
\providecommand{\url}[1]{#1}
\csname url@samestyle\endcsname
\providecommand{\newblock}{\relax}
\providecommand{\bibinfo}[2]{#2}
\providecommand{\BIBentrySTDinterwordspacing}{\spaceskip=0pt\relax}
\providecommand{\BIBentryALTinterwordstretchfactor}{4}
\providecommand{\BIBentryALTinterwordspacing}{\spaceskip=\fontdimen2\font plus
\BIBentryALTinterwordstretchfactor\fontdimen3\font minus
  \fontdimen4\font\relax}
\providecommand{\BIBforeignlanguage}[2]{{%
\expandafter\ifx\csname l@#1\endcsname\relax
\typeout{** WARNING: IEEEtran.bst: No hyphenation pattern has been}%
\typeout{** loaded for the language `#1'. Using the pattern for}%
\typeout{** the default language instead.}%
\else
\language=\csname l@#1\endcsname
\fi
#2}}
\providecommand{\BIBdecl}{\relax}
\BIBdecl

\bibitem{gao2017identifying}
Q.~Gao, F.~Zhou, K.~Zhang, G.~Trajcevski, X.~Luo, and F.~Zhang, ``Identifying
  human mobility via trajectory embeddings.'' in \emph{IJCAI}, vol.~17, 2017,
  pp. 1689--1695.

\bibitem{sun2021trajectory}
T.~Sun, Y.~Xu, F.~Wang, L.~Wu, T.~Qian, and Z.~Shao, ``Trajectory-user link
  with attention recurrent networks,'' in \emph{2020 25th International
  Conference on Pattern Recognition (ICPR)}.\hskip 1em plus 0.5em minus
  0.4em\relax IEEE, 2021, pp. 4589--4596.

\bibitem{zhou2018trajectory}
F.~Zhou, Q.~Gao, G.~Trajcevski, K.~Zhang, T.~Zhong, and F.~Zhang,
  ``Trajectory-user linking via variational autoencoder.'' in \emph{IJCAI},
  2018, pp. 3212--3218.

\bibitem{may2020marc}
L.~May~Petry, C.~Leite Da~Silva, A.~Esuli, C.~Renso, and V.~Bogorny, ``Marc: a
  robust method for multiple-aspect trajectory classification via space, time,
  and semantic embeddings,'' \emph{International Journal of Geographical
  Information Science}, vol.~34, no.~7, pp. 1428--1450, 2020.

\bibitem{miao2020trajectory}
C.~Miao, J.~Wang, H.~Yu, W.~Zhang, and Y.~Qi, ``Trajectory-user linking with
  attentive recurrent network,'' in \emph{Proceedings of the 19th international
  conference on autonomous agents and multiagent systems}, 2020, pp. 878--886.

\bibitem{gao2020adversarial}
Q.~Gao, F.~Zhang, F.~Yao, A.~Li, L.~Mei, and F.~Zhou, ``Adversarial mobility
  learning for human trajectory classification,'' \emph{IEEE Access}, vol.~8,
  pp. 20\,563--20\,576, 2020.

\bibitem{yu2020tulsn}
Y.~Yu, H.~Tang, F.~Wang, L.~Wu, T.~Qian, T.~Sun, and Y.~Xu, ``Tulsn: siamese
  network for trajectory-user linking,'' in \emph{2020 International Joint
  Conference on Neural Networks (IJCNN)}.\hskip 1em plus 0.5em minus
  0.4em\relax IEEE, 2020, pp. 1--8.

\bibitem{chen2022mutual}
W.~Chen, S.~Li, C.~Huang, Y.~Yu, Y.~Jiang, and J.~Dong, ``Mutual distillation
  learning network for trajectory-user linking,'' in \emph{IJCAI}, 2022, pp.
  1973--1979.

\bibitem{rao2020lstm}
J.~Rao, S.~Gao, Y.~Kang, and Q.~Huang, ``Lstm-trajgan: A deep learning approach
  to trajectory privacy protection,'' \emph{arXiv preprint arXiv:2006.10521},
  2020.

\bibitem{zhou2021self}
F.~Zhou, Y.~Dai, Q.~Gao, P.~Wang, and T.~Zhong, ``Self-supervised human
  mobility learning for next location prediction and trajectory
  classification,'' \emph{Knowledge-Based Systems}, vol. 228, p. 107214, 2021.

\bibitem{de2013unique}
Y.-A. De~Montjoye, C.~A. Hidalgo, M.~Verleysen, and V.~D. Blondel, ``Unique in
  the crowd: The privacy bounds of human mobility,'' \emph{Scientific reports},
  vol.~3, no.~1, pp. 1--5, 2013.

\bibitem{brockmann2006scaling}
D.~Brockmann, L.~Hufnagel, and T.~Geisel, ``The scaling laws of human travel,''
  \emph{Nature}, vol. 439, no. 7075, pp. 462--465, 2006.

\bibitem{gonzalez2008understanding}
M.~C. Gonzalez, C.~A. Hidalgo, and A.-L. Barabasi, ``Understanding individual
  human mobility patterns,'' \emph{nature}, vol. 453, no. 7196, pp. 779--782,
  2008.

\bibitem{song2010limits}
C.~Song, Z.~Qu, N.~Blumm, and A.-L. Barab{\'a}si, ``Limits of predictability in
  human mobility,'' \emph{Science}, vol. 327, no. 5968, pp. 1018--1021, 2010.

\bibitem{wang2011human}
D.~Wang, D.~Pedreschi, C.~Song, F.~Giannotti, and A.-L. Barabasi, ``Human
  mobility, social ties, and link prediction,'' in \emph{Proceedings of the
  17th ACM SIGKDD international conference on Knowledge discovery and data
  mining}, 2011, pp. 1100--1108.

\bibitem{sadilek2012far}
A.~Sadilek and J.~Krumm, ``Far out: Predicting long-term human mobility,'' in
  \emph{Proceedings of the AAAI Conference on Artificial Intelligence}, 2012,
  pp. 814--820.

\bibitem{krumme2013predictability}
C.~Krumme, A.~Llorente, M.~Cebrian, A.~Pentland, and E.~Moro, ``The
  predictability of consumer visitation patterns,'' \emph{Scientific reports},
  vol.~3, no.~1, pp. 1--5, 2013.

\bibitem{hasan2013spatiotemporal}
S.~Hasan, C.~M. Schneider, S.~V. Ukkusuri, and M.~C. Gonz{\'a}lez,
  ``Spatiotemporal patterns of urban human mobility,'' \emph{Journal of
  Statistical Physics}, vol. 151, no.~1, pp. 304--318, 2013.

\bibitem{mikolov2013efficient}
T.~Mikolov, K.~Chen, G.~Corrado, and J.~Dean, ``Efficient estimation of word
  representations in vector space,'' \emph{arXiv preprint arXiv:1301.3781},
  2013.

\bibitem{rumelhart1985learning}
D.~E. Rumelhart, G.~E. Hinton, and R.~J. Williams, ``Learning internal
  representations by error propagation,'' California Univ San Diego La Jolla
  Inst for Cognitive Science, Tech. Rep., 1985.

\bibitem{kingma2013auto}
D.~P. Kingma and M.~Welling, ``Auto-encoding variational bayes,'' \emph{arXiv
  preprint arXiv:1312.6114}, 2013.

\bibitem{goodfellow2014generative}
I.~Goodfellow, J.~Pouget-Abadie, M.~Mirza, B.~Xu, D.~Warde-Farley, S.~Ozair,
  A.~Courville, and Y.~Bengio, ``Generative adversarial nets,'' \emph{Advances
  in neural information processing systems}, vol.~27, 2014.

\bibitem{bromley1993signature}
J.~Bromley, I.~Guyon, Y.~LeCun, E.~S{\"a}ckinger, and R.~Shah, ``Signature
  verification using a" siamese" time delay neural network,'' \emph{Advances in
  neural information processing systems}, vol.~6, 1993.

\bibitem{fix1989discriminatory}
E.~Fix and J.~L. Hodges, ``Discriminatory analysis. nonparametric
  discrimination: Consistency properties,'' \emph{International Statistical
  Review/Revue Internationale de Statistique}, vol.~57, no.~3, pp. 238--247,
  1989.

\bibitem{rosenblatt1958perceptron}
F.~Rosenblatt, ``The perceptron: a probabilistic model for information storage
  and organization in the brain.'' \emph{Psychological review}, vol.~65, no.~6,
  p. 386, 1958.

\bibitem{joulin2017efficient}
A.~Joulin, M.~Ciss{\'e}, D.~Grangier, H.~J{\'e}gou \emph{et~al.}, ``Efficient
  softmax approximation for gpus,'' in \emph{International conference on
  machine learning}.\hskip 1em plus 0.5em minus 0.4em\relax PMLR, 2017, pp.
  1302--1310.

\bibitem{cho2011friendship}
E.~Cho, S.~A. Myers, and J.~Leskovec, ``Friendship and mobility: user movement
  in location-based social networks,'' in \emph{Proceedings of the 17th ACM
  SIGKDD international conference on Knowledge discovery and data mining},
  2011, pp. 1082--1090.

\bibitem{yang2016participatory}
D.~Yang, D.~Zhang, and B.~Qu, ``Participatory cultural mapping based on
  collective behavior data in location-based social networks,'' \emph{ACM
  Transactions on Intelligent Systems and Technology (TIST)}, vol.~7, no.~3,
  pp. 1--23, 2016.

\bibitem{weeplaces}
{Liu, Young}, ``Weeplaces dataset,'' \url{https://www.yongliu.org/datasets/},
  2014, accessed: 2022-07-05.

\bibitem{mosteller1968data}
F.~Mosteller and J.~W. Tukey, ``Data analysis, including statistics,''
  \emph{Handbook of social psychology}, vol.~2, pp. 80--203, 1968.

\end{thebibliography}

\end{document}